\documentclass[12pt]{article} 
\usepackage[sectionbib]{natbib}
\usepackage{array,epsfig,fancyheadings,rotating}
\usepackage[]{hyperref}  
\usepackage{color}
\usepackage{sectsty, secdot}
\usepackage{color}
\sectionfont{\fontsize{12}{14pt plus.8pt minus .6pt}\selectfont}
\renewcommand{\theequation}{\thesection\arabic{equation}}
\subsectionfont{\fontsize{12}{14pt plus.8pt minus .6pt}\selectfont}

\usepackage{amsmath}
\usepackage{amssymb}
\usepackage{amsfonts}
\usepackage{multirow}
\usepackage{amsthm}
\usepackage{algorithm}
\usepackage{algorithmic}
\usepackage[T1]{fontenc}
\usepackage{dsfont}

\newtheorem{theorem}{Theorem}
\newtheorem{lemma}{Lemma}

\theoremstyle{definition}

\newtheorem{example}{Example}

\def\MSE {\mbox{MSE}}

\def\Var{\mbox{Var}}

\newcommand{\reva}[1]{{\color{red} #1}}
\renewcommand{\reva}[1]{{#1}}

\begin{document}


\renewcommand{\baselinestretch}{2}

\markright{ \hbox{\footnotesize\rm Statistica Sinica
}\hfill\\[-13pt]
\hbox{\footnotesize\rm
}\hfill }

\markboth{\hfill{\footnotesize\rm LIN WANG} \hfill}
{\hfill {\footnotesize\rm BALANCED SUBSAMPLING} \hfill}

\renewcommand{\thefootnote}{}
$\ $\par


\fontsize{12}{14pt plus.8pt minus .6pt}\selectfont \vspace{0.8pc}
\centerline{\large\bf BALANCED SUBSAMPLING FOR BIG DATA}
\vspace{2pt} 
\centerline{\large\bf WITH CATEGORICAL PREDICTORS}
\vspace{.4cm} 
\centerline{Lin Wang} 
\vspace{.4cm} 
\centerline{\it Department of Statistics, Purdue University}
 \vspace{.55cm} \fontsize{9}{11.5pt plus.8pt minus.6pt}\selectfont


\begin{quotation}
\noindent {\it Abstract:}
Supervised learning under measurement constraints is a common challenge in statistical and machine learning. In many applications, despite extensive design points, acquiring responses for all points is often impractical due to resource limitations. Subsampling algorithms offer a solution by selecting a subset from the design points for observing the response. Existing subsampling methods primarily assume numerical predictors, neglecting the prevalent occurrence of big data with categorical predictors across various disciplines. This paper proposes a novel balanced subsampling approach tailored for data with categorical predictors. A balanced subsample significantly reduces the cost of observing the response and possesses three desired merits. First, it is nonsingular and, therefore, allows linear regression with all dummy variables encoded from categorical predictors. Second, it offers optimal parameter estimation by minimizing the generalized variance of the estimated parameters. Third, it allows robust prediction in the sense of minimizing the worst-case prediction error. We demonstrate the superiority of balanced subsampling over existing methods through extensive simulation studies and a real-world application.

\vspace{9pt}
\par
\end{quotation}\par

\def\thefigure{\arabic{figure}}
\def\thetable{\arabic{table}}

\renewcommand{\theequation}{\thesection.\arabic{equation}}

\fontsize{12}{14pt plus.8pt minus .6pt}\selectfont

\section{Introduction}
Supervised learning under measurement constraints is a common challenge in statistical and machine learning \citep{wang2017computationally,meng2021lowcon}. In many applications, despite the availability of extensive predictor observations (design points), acquiring the observations of the response variable for all design points is frequently impractical due to resource limitations. For example, consider a scenario in healthcare where researchers aim to develop a predictive model for patient outcomes based on a diverse set of health-related predictors. A large set of predictor observations, such as patient demographics, medical history, and genetic information, is readily available. However, obtaining the corresponding response variable, such as a medical condition or treatment outcome, may involve invasive procedures or expensive diagnostic tests. Given the constraints of limited resources, observing the response of every individual in the dataset becomes practically impossible. Consequently, selecting an informative subsample from the set of design points to observe becomes crucial and challenging. 

In recent years, there has been a growing interest in developing design-based optimal subsampling methods.
Most existing methods focus on numerical predictors in various learning models, such as linear regression \citep{wang2019information,ma2015statistical,wang2021orthogonal}, generalized linear models \citep{wang2018optimalsubsampling,ai2021optimal2,cheng2020information}, linear mixed models \citep{zhu2024group}, quantile regression \citep{wang2021optimal,ai2021optimal1}, nonparametric regression \citep{meng2020more,zhang2024independence}, Gaussian process modeling \citep{he2022gaussian}, and model-free scenarios \citep{mak2018support,shi2021model,song2022scsampler}.

Big data with categorical predictors are frequently encountered in many scientific research areas \citep{huang2014feature,zuccolotto2018big,johnson2018studying}. Numerical predictors may also be binned into categorical ones for better modeling and interpretation \citep{kanda2013investigation,yu2022distributed}. 
Despite numerous studies on subsampling methods, they do not apply to data with categorical predictors, so researchers have no choice but to use simple random subsamples, for example, see \cite{maronna2000robust} and \cite{yu2022distributed}. However, simple random subsamples may bring significant issues, especially for data with categorical predictors. To illustrate, consider that categorical predictors are commonly encoded using dummy variables in regression models. In a dataset with $p$ categorical predictors, each having $q_j$ levels, $j=1,\ldots,p$, these predictors are coded to $\sum_{j=1}^{p}(q_j-1)$ dummy variables, which substantially increases the dimensionality of the regression task. Thus, as pointed out by \cite{maronna2000robust} and \cite{koller2017nonsingular}, singular subsamples (or more accurately, subsamples with singular information matrices) are frequently encountered when dealing with categorical predictors due to the high dimensionality of the dummy variables. Consequently, a simple random subsample cannot facilitate the estimation of the effect for every dummy variable, even though the full data allows for such estimation. This deficiency arises because the subsample lacks crucial information in the full data, and avoiding such substantial information loss is paramount. Furthermore, even among nonsingular subsamples, there can be significant variations in the accuracy of parameter estimation. Identifying the subsample that enables optimal parameter estimation is also important.

This paper proposes a novel method named ``balanced subsampling'' designed specifically for subsampling data with categorical predictors.
The selected subsample achieves a combinatorial balance between values (levels) of the predictors and, therefore, enjoys three desired merits. First, a balanced subsample is generally nonsingular and thus allows the estimation of all parameters in linear (ANOVA) regression.
Second, a balanced subsample provides the optimal parameter estimation in the sense of minimizing the generalized variance of the estimated parameters. Third, when the established model is used for prediction, the model trained on a balanced subsample provides robust predictions in the sense of minimizing the worst-case prediction error.
For practical use, we develop an algorithm that sequentially selects data points from the full data to obtain a balanced subsample.

The remainder of the paper is organized as follows.
Section 2 presents the issues of simple random subsampling for data with categorical predictors, which motivates us to develop a new subsampling method. Section 3 proposes the balanced subsampling method and develops an efficient algorithm for sequentially subsampling from big data.
Section 4 examines the performance of balanced subsampling through extensive simulations, and Section 5 demonstrates the utility of using balanced subsamples in a real-world application.
Section 6 offers concluding remarks. Supplementary Materials provide proof of technical results and discuss the computational complexity of the proposed algorithm.

\section{Motivations}
Let $X=(x_1,\ldots,x_N)^T$ denote the design matrix of the full data, where $x_i=(x_{i1},\ldots,x_{ip})^T$ consists of the values of $p$ categorical predictors, each with $q_j$ levels for $j=1,\ldots,p$ and is coded to $q_j-1$ binary dummy variables. 
Let $z_{i,jk}$ be the value of the $k$th dummy variable for $x_{ij}$ and $Z$ the matrix formed by $z_{i,jk}$. 
Linear regression on the dummy variables is given by:
\begin{equation}\label{slm}
y_i=\beta_0+\sum_{k=1}^{q_1-1}\beta_{1k} z_{i,1k}+\cdots+\sum_{k=1}^{q_p-1}\beta_{pk} z_{i,pk} + \varepsilon_i,
\end{equation}
where $\beta_{jk}$ are parameters to be estimated and $\varepsilon_i$ is the independent random error with mean 0 and variance $\sigma^2$. It is intuitive to assume that $M=Z^TZ$ is nonsingular, enabling linear regression on the full data if all responses can be observed.

We consider taking a subsample of size $n$ from the full data $X$, denoted as $X_s$, and observe its corresponding response vector $y_s$. 
The OLS estimator for $\beta=(\beta_0,\beta_{11},\ldots,\beta_{p(q_p-1)})^T$ based on the subsample is given by
\begin{equation}\label{betas}
\hat{\beta}_s = (Z_s^TZ_s)^{-1}(Z_s^Ty_s),
\end{equation}
where $Z_s=(z_{1}^*,\ldots,z_{n}^*)^T$ are the rows in $Z$ corresponding to points in $X_s$.
We have three concerns regarding the subsample $X_s$.

\subsection{Nonsingularity}
The subsample should allow the estimation of all parameters in $\beta$, which is possible only if the information matrix, $M_s=Z_s^TZ_s$, is nonsingular. \reva{However, singular subsamples (subsamples with a singular information matrix) are frequently encountered when dealing with categorical predictors,} which can be illustrated by the following two toy examples.
\begin{example}\label{exa1}
Assume the full data contain a single categorical predictor with 2 repetitions of 5 levels, that is, $X=(1,1,2,2,3,3,4,4,5,5)^T$. Use dummy variables, then
$$
Z=\left(
    \begin{array}{cccccccccc}
      1 & 1 & 1 & 1 & 1 & 1 & 1 & 1 & 1 & 1 \\
      1 & 1 & 0 & 0 & 0 & 0 & 0 & 0 & 0 & 0 \\
      0 & 0 & 1 & 1 & 0 & 0 & 0 & 0 & 0 & 0 \\
      0 & 0 & 0 & 0 & 1 & 1 & 0 & 0 & 0 & 0 \\
      0 & 0 & 0 & 0 & 0 & 0 & 1 & 1 & 0 & 0 \\
    \end{array}
  \right)^T.
$$
Consider choosing a subsample $X_s$ with 5 points, then $M_s$ is nonsingular only if $X_s$ contains at least one observation of each level. Out of the $\binom{10}{5}=252$ possible subsamples, only $2^5$ of them are good in this way, and the probability of obtaining such a subsample with simple random sampling is $2^5/252=12.7\%$.
\end{example}

\begin{example}\label{exa2}
Suppose the full data have $N=1000$ points and $p=2$ predictors. We generate data from an independent bivariate normal distribution with mean 0 and variance 1, divide the range of either predictor into 5 equal-sized intervals, and code the values according to which interval they fall. Then each predictor includes 5 levels, and the two predictors have 25 possible pairs of levels. 
We select a subsample of size $n=25$. There are $\binom{1000}{25}\approx 10^{49}$ possible subsamples from simple random sampling. An exhaustive examination of all those subsamples is infeasible. Therefore, we randomly investigate $10^5$ of them, and only 4.81\% have nonsingular information matrices. It is not easy to obtain a nonsingular subsample from simple random sampling.
\end{example}

\subsection{Optimal estimation}
Even among the nonsingular subsamples, the accuracy of parameter estimation varies greatly across different subsamples.
\begin{example}\label{exa3}
We continue Example \ref{exa2}. For all the nonsingular subsamples with $n=25$ (out of the $10^5$ investigated random subsamples), we generate the response variable $y$ through the model
\begin{equation}\label{eq3}
y_i = 1 + z_{i,11}+\cdots+ z_{i,14} + z_{i,21} + \cdots + z_{i,24} + \varepsilon_i,
\end{equation}
where $\varepsilon_i\sim N(0,1)$, and train the model in \eqref{slm} on each of the nonsingular subsamples. We repeat this process $T=1000$ times and examine the empirical mean squared error (MSE):
\begin{equation}\label{c4mse}
\MSE=T^{-1}\sum_{t=1}^{T}\|\hat{\beta}^{(t)}-\beta\|^2,
\end{equation}
where $\hat{\beta}^{(t)}$ is the OLS estimate of $\beta$ via a subsample in the $t$th repetition, $t=1,\ldots,1000$.
Figure \ref{hist3} (left) shows the histogram of $\log_{10}$(MSE) for all nonsingular subsamples. The MSE varies dramatically, with the minimum as low as $10^{0.85}=7.1$ achieved by only a couple of subsamples. 
Recall that we examined $10^5$ random subsamples, and only a couple of them allows the ``optimal'' estimation. It is very hard to obtain such ``optimal'' subsamples from simple random sampling.
\begin{figure}[t]
  \centering
  \includegraphics[width=.8\textwidth]{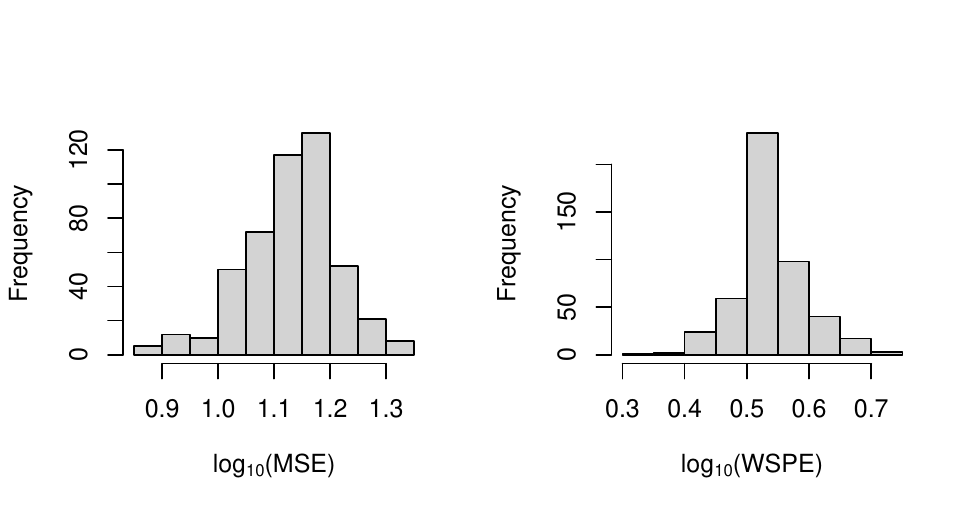}
  \vspace{-1cm}
  \caption{Histograms of $\log_{10}$(MSE) and $\log_{10}$(WSPE) of the trained model on each subsample with a nonsingular information.}\label{hist3}
\end{figure}
\end{example}

\subsection{Robust prediction}
We hope the trained model on a subsample provides ``robust'' prediction, where the terminology ``robust'' can be understood in the sense of performing well in the worst-case scenario. 

\begin{example}\label{exa4}
We continue Example \ref{exa3} and examine the empirical worst-case squared prediction error (WSPE) for all nonsingular subsamples:
\begin{equation}\label{wmspe}
\mbox{WSPE}=\max_{x\in \mathcal{X}} \left\{T^{-1}\sum_{t=1}^{T}(y^{(t)}-z^T\hat{\beta}^{(t)})^2\right\}
\end{equation}
where $\mathcal{X}$ includes all the 25 possible pairs of levels for the two predictors, $z$ is the vector of dummy variables for any $x\in \mathcal{X}$, $y^{(t)}$ is the response in the $t$th repetition, and $\hat{\beta}^{(t)}$ is the OLS estimate of $\beta$ via a subsample, for $t=1,\ldots,1000$. Figure \ref{hist3} (right) shows the histogram of $\log_{10}$(WSPE) for all the nonsingular subsamples.
The minimum of WSPE is $10^{0.31}=2.0$ achieved by a single subsample, which is, again, almost impossible to obtain from simple random sampling.
\end{example}

\section{Balanced Subsampling}
In this section, we propose the balanced subsampling method and develop a computationally efficient algorithm to implement it. The proposed method targets the above three concerns: providing a nonsingular subsample, enabling optimal parameter estimation, and ensuring robust prediction.
\subsection{The method}
We first consider the nonsingularity of a subsample and provide the following important result.
\begin{theorem}\label{th1}
Let $\lambda_{\min}(M_s)$ be the smallest eigenvalue of $M_s$. For a subsample $X_s$,
$$
\lambda_{\min}(M_s)\geq n\nu(1-f(X_s))
$$
where $n$ is the subsample size of $X_s$,
$\nu$ is a positive constant independent of $X_s$,
\begin{equation}\label{fz}
f(X_s)=\sqrt{\sum_{j=1}^{p}\sum_{u=1}^{q_{j}}q_{j}^2\left[\frac{1}{q_{j}}-\frac{n_{j}(u)}{n}\right]^2
+\sum_{j=1}^{p}\sum_{k=1,k\neq j}^{p}\sum_{u=1}^{q_{j}}\sum_{v=1}^{q_{k}}q_jq_k\left[\frac{1}{q_jq_k}-\frac{n_{jk}(u,v)}{n}\right]^2},
\end{equation}
$n_j(u)$ is the number of times that the $u$th level of the $j$th predictor is observed in $X_s$, and $n_{jk}(u,v)$ is the number of times that the pair of levels $(u,v)$ is observed for the $j$th and $k$th predictors in $X_s$. Therefore, $M_s$ is nonsingular if $f(X_s)<1$.
\end{theorem}

Theorem \ref{th1} indicates that we can search for the subsample that minimizes $f(X_s)$ to ensure that the $M_s$ is nonsingular.
By \eqref{fz}, $f(X_s)$ has two critical components: (a) $f_j=\sum_{u=1}^{q_j}[1/q_j-n_j(u)/n]^2$ that measures the balance of levels for the $j$th predictor, and (b) $f_{jk}=\sum_{u=1}^{q_j}\sum_{v=1}^{q_k}[1/(q_jq_k)-n_{jk}(u,v)/n]^2$ that measures the balance of level combinations for the $j$th and $k$th predictors. Clearly, if $f_j=0$, all levels of the $j$th predictor are observed the same number of times in $X_s$ so that they achieve the perfect balance; if $f_{jk}=0$, all pairs of levels for the $j$th and $k$th predictors are observed the same number of times in $X_s$. Such balance is called combinatorial orthogonality, and a matrix possessing combinatorial orthogonality is called an orthogonal array.

Generally,
an orthogonal array of strength $t$ is a matrix where entries of each column of the matrix come from a fixed finite set of $q_j$ levels for $j=1,\ldots,p$, arranged in such a way that all ordered $t$-tuples of levels appear equally often in every selection of $t$ columns of the matrix. The $t$ is called the strength of the orthogonal array. Readers are referred to \cite{hedayatorthogonal} for a comprehensive introduction to orthogonal arrays.
Here is an example of an orthogonal array with $p=3$ predictors, each having 3 levels, and strength $t=2$:
$$
\left(
  \begin{array}{ccccccccc}
1 & 1 & 1 & 2 & 2 & 2 & 3 & 3 & 3 \\
1 & 2 & 3 & 1 & 2 & 3 & 1 & 2 & 3\\
1 & 2 & 3 & 2 & 3 & 1 & 3 & 1 & 2\\
  \end{array}
\right)^T.
$$
Each pair of levels in any two columns of the orthogonal array appears once. Clearly, we have the following lemma.

\begin{lemma}\label{lem2}
A subsample $X_s$ forms an orthogonal array of strength two if and only if $f(X_s)=0$.
\end{lemma}

Now we show that the subsample minimizing $f(X_s)$ also allows the optimal estimation of parameters. To see this, note that
$\hat{\beta}_s$ in \eqref{betas} is an unbiased estimator of $\beta$ with
\begin{equation}\label{ms}
\Var(\hat{\beta}_s|X_s)=\sigma^2M_s^{-1} = \sigma^2(Z_s^TZ_s)^{-1}.
\end{equation}
The $\Var(\hat{\beta}_s|X_s)$ is a function of $X_s$ (in the form of $Z_s$), which indicates again that the subsampling strategy is critical in reducing the variance of $\hat{\beta}_s$.
To minimize $\Var(\hat{\beta}_s|X_s)$, we seek the $X_s$ which, in some sense, minimizes $M_s^{-1}$. This is typically done, in experimental design strategy, by minimizing an optimality function $\psi(M_s^{-1})$ of the matrix $M_s^{-1}$ \citep{kiefer1959optimum1,atkinson2007optimum}.
A common choice for $\psi$ is the determinant, which is akin to the criterion of $D$-optimality for the selection of optimal experimental designs.

\begin{theorem}\label{thopt}
A subsample $X_s$ is $D$-optimal for the model in \eqref{slm} if $f(X_s)=0$.
\end{theorem}

\cite{cheng1980orthogonal} showed that an orthogonal array of strength two is universally optimal, i.e., optimal under a wide variety of criteria by minimizing the sum of a convex function of coefficient matrices for the reduced normal equations. However, Cheng's result does not apply to the dummy coding system, so that his result does not include Theorem \ref{thopt}. To the best of our knowledge, Theorem \ref{thopt} originally shows the optimality of orthogonal arrays for the commonly used dummy coding for categorical predictors.

Next, we show that minimizing $f(X_s)$ also allows robust prediction. 
Let $\mathcal{X}$ denote the set of all possible level combinations of predictors, that is, $\mathcal{X}=\{x=(x_1,\ldots,x_p): x_j=1,\ldots,q_j, j=1,\ldots, p\}$, then $\#\mathcal{X}=\prod_{j=1}^{p}q_j$. For any $x\in \mathcal{X}$, let $z$ be the coded vector of $x$ and $Y$ the random variable of its response with $E(Y)=z\beta$ and $\Var(Y)=\sigma^2$. The WSPE is given by $\max_{x\in \mathcal{X}}E[(Y-z^T\hat{\beta}_s)^2|X_s]$, where
\begin{equation}\label{minm}
    E[(Y-z^T\hat{\beta}_s)^2|X_s]=E[(Y-z^T\beta)^2]+E[(z^T\beta-z^T\hat{\beta}_s)^2|X_s]=\sigma^2(1+z^TM_s^{-1}z).
\end{equation}
The WSPE is a function of $X_s$ (in the form of $M_s$), which indicates again that the subsampling strategy is critical in reducing WSPE.
The following theorem shows that the WSPE is minimized when $f(X_s)=0$.
\begin{theorem}\label{thminmax}
Let $Q=1+\sum_{j=1}^{p}(q_j-1)$. For a subsample $X_s$ of size $n$,
\begin{equation}\label{lbwspe}
\max_{x\in \mathcal{X}}E[(Y-z^T\hat{\beta}_s)^2|X_s]\geq \sigma^2(1+Q/n).
\end{equation}
The equality in \eqref{lbwspe} holds if $f(X_s)=0$.
\end{theorem}

Theorems \ref{th1}--\ref{thminmax} indicate that $f(X_s)=0$ ensures model estimability as well as optimal estimation and robust prediction. Considering that a full dataset may generally do not contain a subset with exact zero $f(X_s)$, the objective of the balanced subsampling is to achieve an approximate balance via the optimization problem:
\begin{eqnarray}\label{optfz}
  X_s^*&=&\arg\min_{X_s\subseteq X}f(X_s) \\
   &s.t.& X_s \mbox{ contains } n \mbox{ points.}\nonumber
\end{eqnarray}

The optimization problem in \eqref{optfz} is not easy to solve. The computation of $f(X_s)$ requires the examination of balance for every single predictor and every pair of predictors in $X_s$, so it requires $O(np^2)$ operations to compute $f(X_s)$ for any $X_s$. In addition,
an exhaustive search for all possible $X_s$ requires $O(N^n)$ operations, making it infeasible for even moderate sizes of the full data.
There are many types of algorithms for finding optimal designs and among them, exchange algorithms are among the most popular.
For the reasons argued in \cite{wang2021orthogonal}, these algorithms are cumbersome in solving the subsampling problem in \eqref{optfz}. We will propose a sequential selection algorithm to efficiently select subsample points.

\subsection{A sequential algorithm}
The following result is critical in developing the algorithm.
\begin{theorem}\label{th4}
For a subsample $X_s=(x_{ij}^*)$, $i=1,\ldots,n$ and $j=1,\ldots,p$,
\begin{equation}\label{delt}
f^2(X_s)=2n^{-2}\sum_{1\leq i<l\leq n}[\delta(x_i^*,x_l^*)]^2+C,
\end{equation}
where
\begin{equation}\label{delta}
\delta(x_i^*,x_l^*)= \sum_{j=1}^{p}q_j\delta_1(x_{ij}^*,x_{lj}^*),
\end{equation}
$\delta_1(x_{ij}^*,x_{lj}^*)$ is 1 if $x_{ij}^*=x_{lj}^*$ and 0 otherwise, and $C=n^{-1}(\sum_{j=1}^{p}q_j)^2+p-\sum_{j=1}^{p}q_j-p^2$.
\end{theorem}

By Theorem \ref{th4}, the optimization in \eqref{optfz} can be achieved by minimizing $\sum_{1\leq i<l\leq n}[\delta(x_i^*,x_l^*)]^2$.
To select an $n$-point subsample, we start with a random point $x_1^*$ and select $x_2^*,\ldots,x_n^*$ sequentially. Suppose we have already selected $m$ points,
then the $(m+1)$th point is selected by
$$
x_{m+1}^* = \arg\min_{x} ~\left\{\sum_{i=1}^{m-1}\sum_{l=i+1}^{m}[\delta(x_i^*,x_l^*)]^2 + \sum_{i=1}^{m}[\delta(x_i^*,x)]^2 \right\}\nonumber\\
= \arg\min_{x} \Delta(x)
$$
where
\begin{equation}\label{eq4}
\Delta(x)= \sum_{i=1}^{m}[\delta(x_i^*,x)]^2,
\end{equation}
and the minimization is over $x\in X\backslash\{x_1^*,\ldots,x_m^*\}$.
Since
$$
\Delta(x)= \sum_{i=1}^{m-1}[\delta(x_i^*,x)]^2+[\delta(x_m^*,x)]^2,
$$
we only need to compute $\delta(x_m^*,x)$ to update $\Delta(x)$ in the $(m+1)$th iteration. \reva{Each iteration has a complexity of $O(Np)$, and the overall complexity of the algorithm is $O(Npn)$.} Algorithm \ref{alg1} outlines this sequential selection algorithm. 

\begin{algorithm}[t!]
\caption{Balanced Subsampling (Sequential Selection)}
\label{alg1}
\begin{algorithmic}
\STATE \textbf{Input:} a sample (dataset) $X$, a required subsample size $n$\\
\textbf{Output:} a subsample $X_s$
\STATE Set $m=1$ and randomly select $x_1^*$ from $X$
\FOR{each $x\in X\backslash\{x_1^*\}$}
\STATE Compute $\Delta(x)$ via \eqref{eq4}
\ENDFOR
\WHILE{$m< n $}
\STATE Find $x_{m+1}^*=\arg\min_x \Delta(x)$ and include $x_{m+1}^*$ in $X_s$
\FOR{each $x\in X\backslash\{x_1^*,\ldots,x_{m+1}^*\}$}
\STATE Update $\Delta(x)\leftarrow\Delta(x)+[\delta(x_{m+1}^*,x)]^2$
\ENDFOR
\STATE $m \leftarrow m + 1$
\ENDWHILE
\end{algorithmic}
\end{algorithm}

\begin{figure}[h]
    \centering
    \includegraphics[width=\linewidth]{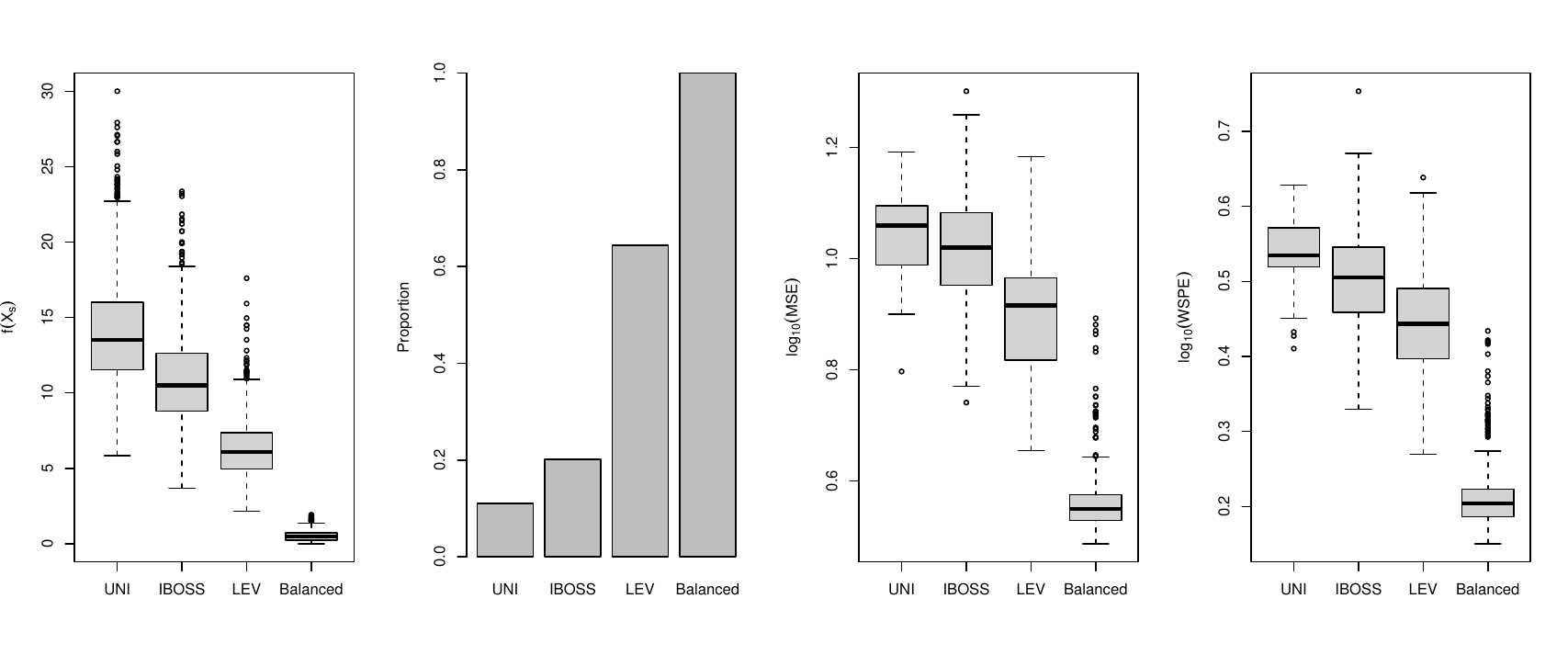}
    \vspace{-1.8cm}
    \caption{The values of $f(X_s)$, proportion of nonsingular subsamples, MSE, and WSPE for subsamples generated using various methods.}
    \label{exp5box}
\end{figure}

\begin{figure}[h]
  \centering
  \includegraphics[width=.9\textwidth]{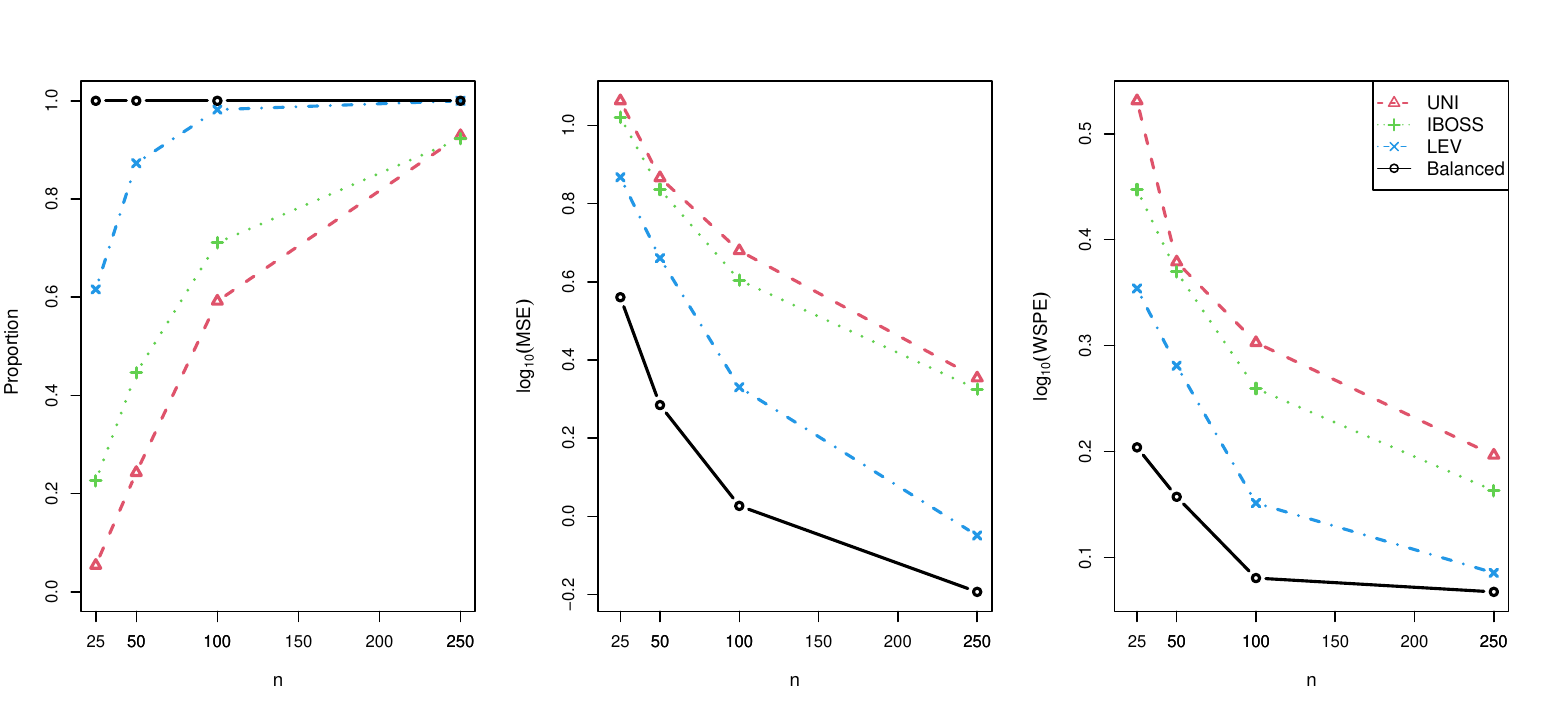}
  \vspace{-.8cm}
  \caption{The proportions of nonsingular subsamples (left), MSEs of estimated parameters (middle), and WSPEs (right) for different subsampling methods in Example \ref{toy}.}\label{toymse}
\end{figure}

\begin{example}\label{toy}\rm
We continue Examples \ref{exa2}--\ref{exa4} to demonstrate the effectiveness of Algorithm \ref{alg1} by comparing it with existing popular subsampling methods, including simple random sampling (denoted as UNI for consistency with the literature, as it assigns a uniform weight to all observations), information-based optimal subdata selection (IBOSS, \cite{wang2019information}), and leveraging subsampling (LEV, \cite{ma2015leveraging}). We generate the full data following the procedure outlined in Example \ref{exa2} and select subsamples of size $n=25$ using different methods. We repeat this process 1000 times. The left two panels of Figure \ref{exp5box} plot the values of $f(X_s)$ and proportions of nonsingular subsamples for each method. UNI often misses many pairs of levels, resulting in high $f(X_s)$ values and a low proportion of nonsingular subsamples. IBOSS and LEV, applied to the dummy variables, offer better balance than UNI, resulting in higher proportions of nonsingular subsamples. Balanced subsamples obtained from Algorithm \ref{alg1} consistently exhibit smaller $f(X_s)$ values, indicating a higher degree of balance compared to other subsamples, and are consistently nonsingular. For each subsample, we repeatedly generate the response for $T=1000$ times through the model in \eqref{eq3} and examine the empirical MSE in \eqref{c4mse} and WSPE in \eqref{wmspe}, displayed in the right two panels of Figure \ref{exp5box}. The balanced subsamples demonstrate significantly smaller MSEs and WSPEs, clearly outperforming other subsampling methods.

We further explore the performance of subsampling methods across different subsample sizes $n=25,50,100,250$. \reva{For each subsample size, we repeatedly generate the full data and the response for $T=1000$ times and select a subsample of size $n$ using different methods. Figure \ref{toymse} plots the proportions of nonsingular subsamples, the empirical MSE \eqref{c4mse}, and the empirical WSPE \eqref{wmspe}.} The balanced subsamples always significantly outperform other approaches for any subsample size due to their balance on levels of predictors. Specifically, a balanced subsample is consistently nonsingular. Increasing the subsample size may enhance the nonsingularity of other subsamples, but they still provide much worse parameter estimation and response prediction than a balanced subsample. 
\end{example}

\section{Simulation studies}
We conduct simulation studies to assess the merits of the balanced subsampling method relative to existing subsampling schemes. Consider $p=20$ predictors each with $q_j=j+1$ levels for $j=1,\ldots,p$.
The simulation is replicated $T=1000$ times.
In each replication, we generate values of the predictors under three structures:
\begin{itemize}
  \item[Case 1.] Covariates are independent, and each follows a discrete uniform distribution with $q_j$ levels.
  \item[Case 2.] Covariates are independent, and for each predictor, the $q_j$ levels have probabilities proportional to $1,\ldots,q_j$.
  \item[Case 3.] Generate each point $x_i$ from multivariate normal distribution: $x_i\sim N(0,\Sigma)$ with
  \begin{equation}\label{Sigm}
  \Sigma=\left(0.5^{\xi(j,k)}\right),
  \end{equation}
  where $\xi(j,k)$ is equal to 0 if $j=k$ and 1 otherwise. Discretize $[-3,3]$ to $q_j$ intervals of equal length, and let $x_{ij}=u$ if $x_{ij}$ falls into the $u$th interval. Let $x_{ij}=1$ if $x_{ij}<-3$ and $x_{ij}=q_j$ if $x_{ij}>3$.
\end{itemize}
The response data are generated from the linear model in \eqref{slm} with the true value of $\beta$ being a vector of unity and $\sigma=1$. 
We investigate four settings of the full data size $N=5\times10^3, 10^4, 5\times 10^4$, and $10^5$, and two settings of the subsample size $n=500$ and 2000. Four subsampling approaches, UNI, IBOSS, LEV, and the balanced subsampling, are evaluated by comparing the proportions of nonsingular subsamples, MSEs \eqref{c4mse}, and WSPEs \eqref{wmspe}. \reva{To accelerate LEV, we use a fast Singular Value Decomposition method implemented in the R package ``corpcor''. Since the set of all level combinations $\mathcal{X}$ contains $\prod_{j=1}^p q_j=5\times 10^{19}$ points, it is infeasible to evaluate predictive performance across the entire set $\mathcal{X}$. Instead, we randomly sample $10^6$ points in $\mathcal{X}$ to compute WSPE. Note that the comparisons of subsampling approaches on MSE and WSPE are independent of the settings of the true parameters $\beta$ or $\sigma$, as can be seen from \eqref{ms} and \eqref{minm}.}
\begin{figure}[t]
  \centering
  \includegraphics[width=.75\textwidth]{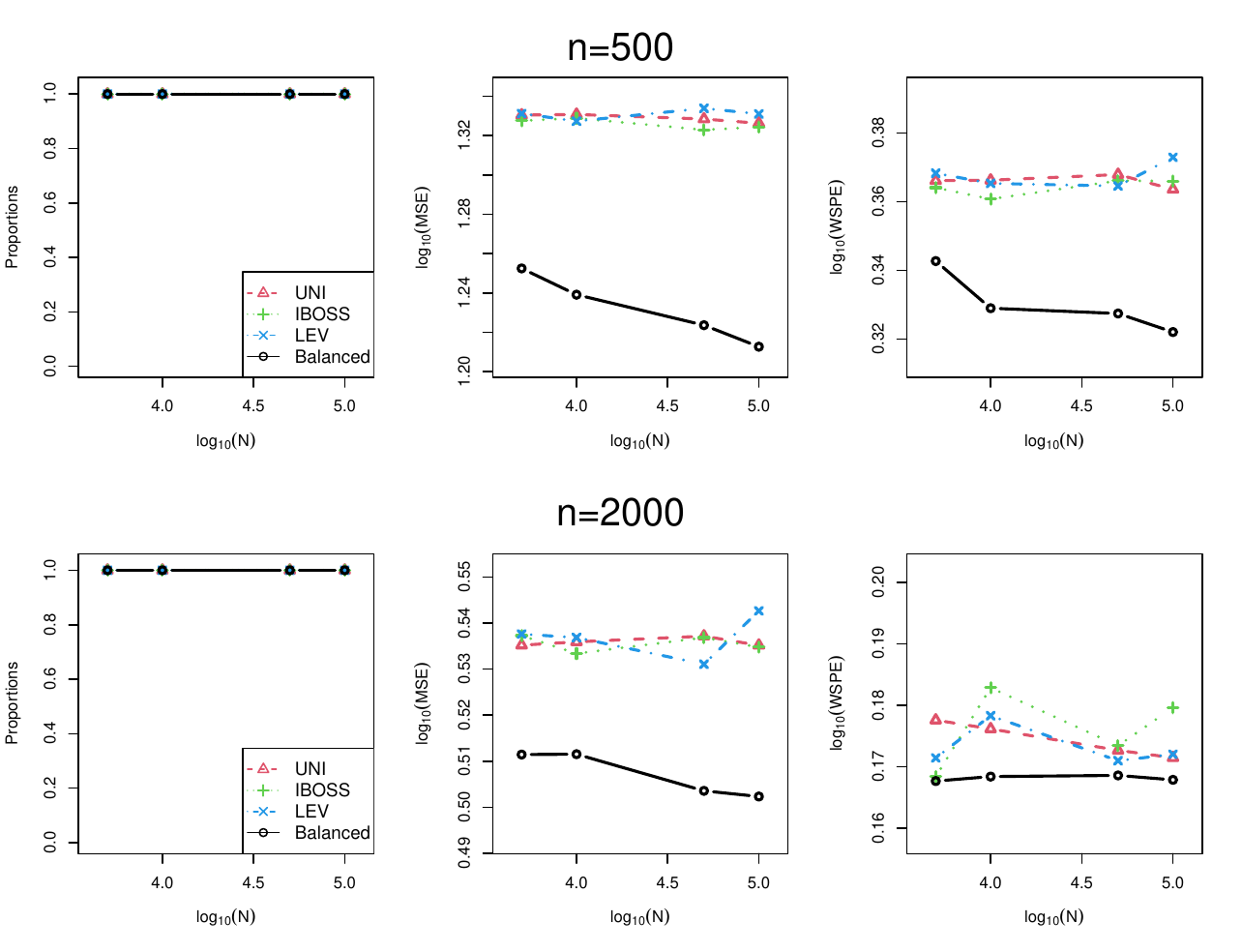}
  \vspace{-6mm}
  \caption{The proportions of nonsingular subsamples (left), MSEs of estimated parameters (middle), and WSPEs (right) for different subsampling methods for predictors in Case 1.}\label{sim3}
\end{figure}

Figure \ref{sim3} compares the subsampling methods for the full data generated in Case 1. In this case, predictor levels in the full data are highly balanced, so all subsamples closely resemble balance and are nonsingular. Even so, the balanced subsamples consistently provide more accurate parameter estimation and slightly better prediction than other methods. 
Note that when the full data are highly balanced, the increase in the full data size does not have a substantial contribution to the balance of subsamples. Therefore, balanced subsampling exhibits only a slight improvement in parameter estimation and relatively flat WSPE as the full data size increases. Considering that we can only examine a subset of $\mathcal{X}$, the WSPE may even slightly fluctuate as the full data size increases.

\begin{figure}[t]
  \centering
  \includegraphics[width=.75\textwidth]{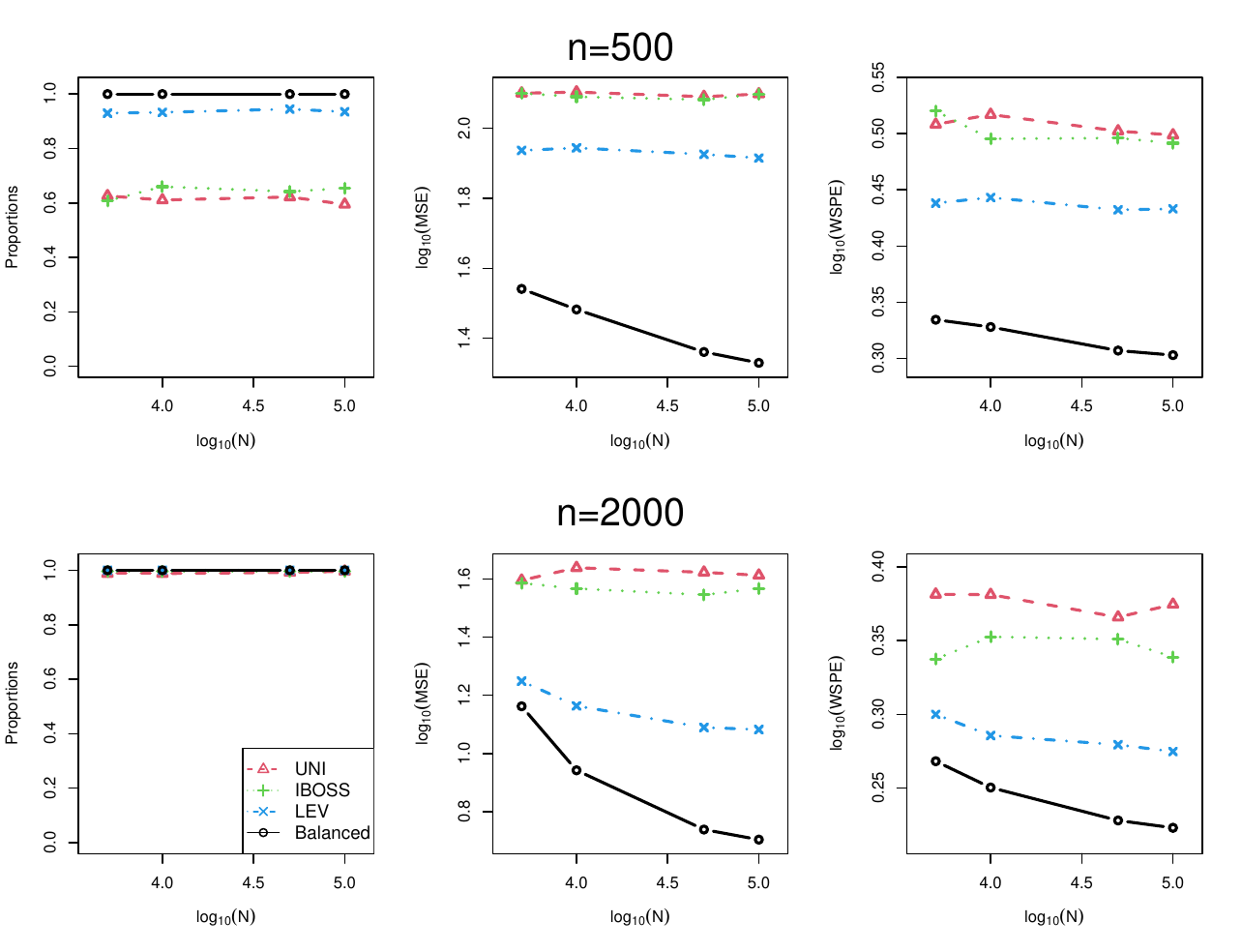}
  \vspace{-6mm}
  \caption{The proportions of nonsingular subsamples (left), MSEs of estimated parameters (middle), and WSPEs (right) for different subsampling methods for predictors in Case 2.}\label{sim2}
\end{figure}

Figure \ref{sim2} plots results for Case 2 where levels of predictors are unbalanced, which is typically the case in real practice. We observe that, when the subsample size is $n=500$, only around 60\% of UNI and IBOSS subsamples are nonsingular, 90\% of LEV subsamples are nonsingular, whereas all balanced subsamples are nonsingular. With just 500 observations, it becomes possible to estimate a model that includes all dummy variables, indicating that a balanced subsample enables significant cost savings in observing the response. 
Increasing the subsample size to $n=2000$ may improve the proportions of nonsingularity for other methods, but the estimation and prediction obtained from those subsamples are still much worse than the balanced subsamples. Notably, a balanced subsample with $n=500$ exhibits better accuracy in parameter estimation when compared with UNI and IBOSS subsamples with $n=2000$ (around 1.6 on $\log_{10}(\MSE)$). This observation once again underscores the significant savings achieved by utilizing a balanced subsample.
More importantly, the MSEs and WSPEs from the balanced subsamples decrease fast as the full data size $N$ increases, even though the subsample size is fixed at $n=500$ or 2000. This trend demonstrates that the balanced subsampling extracts more information from the full data with a fixed subsample size when the full data are more informative. IBOSS has this nice property for continuous predictors \citep{wang2019information}, but not for categorical predictors because of the high association between dummy variables.

\begin{figure}[h]
  \centering
  \includegraphics[width=.75\textwidth]{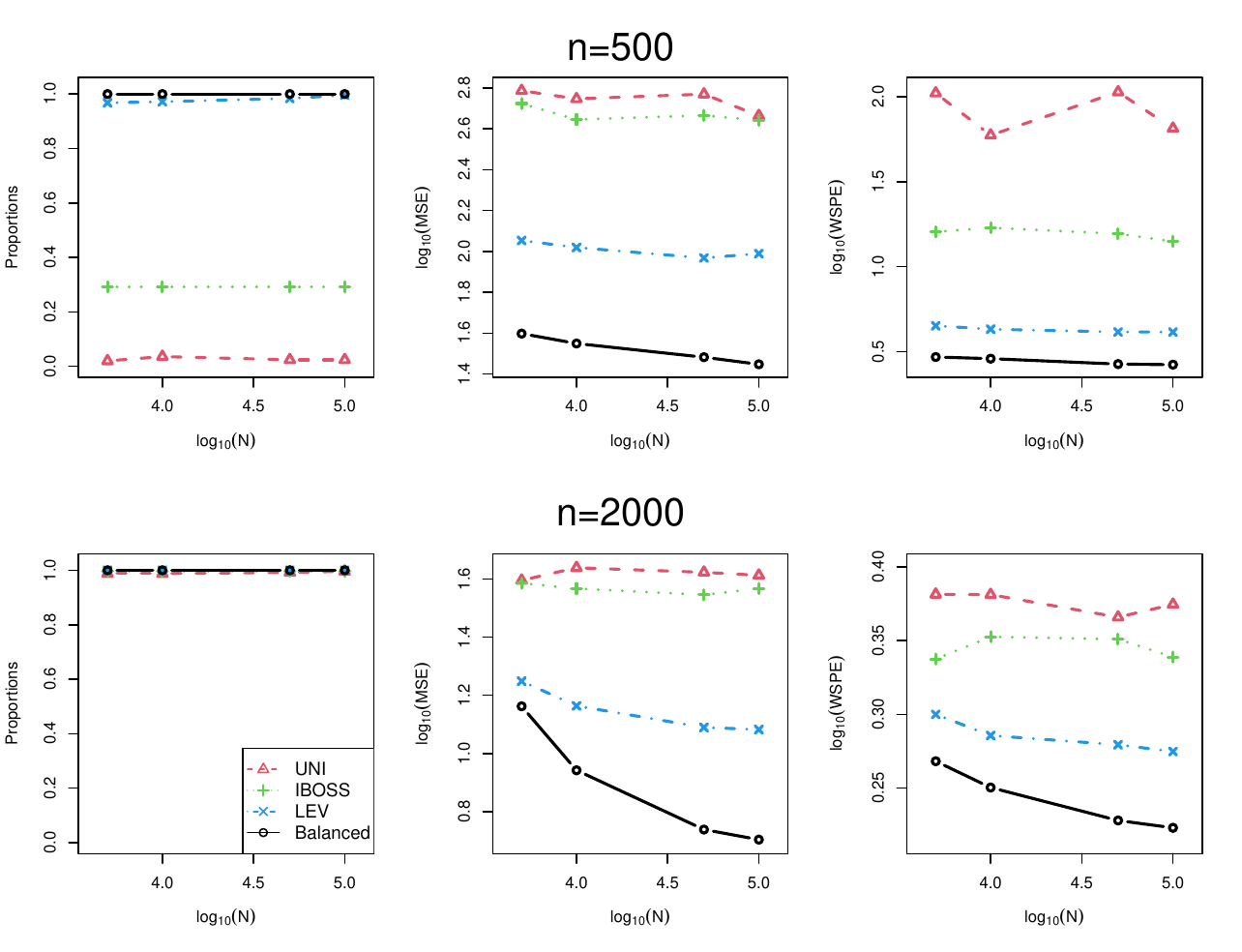}
  \vspace{-6mm}
  \caption{The proportions of nonsingular subsamples (left), MSEs of estimated parameters (middle), and WSPEs (right) for different subsampling methods for predictors in Case 3. }\label{sim1}
\end{figure}

Figure \ref{sim1} examines Case 3 where predictors are correlated in the full data. For $n=500$, almost all UNI subsamples are singular, and only less than 40\% of IBOSS subsamples are nonsingular. LEV performs well in terms of nonsingularity but has worse MSE and WSPE compared to balanced subsampling. For either setting of the subsample size, we observe a greater superiority of the balanced subsamples and a decreasing trend of MSEs and WSPEs as the full data size increases. This is because the balanced subsamples have reduced correlation and enhanced combinatorial orthogonality between predictors, which helps reduce the collinearity between predictors and therefore allows a more accurate estimate of parameters. Notably, a balanced subsample with $n=500$ exhibits comparable accuracy in parameter estimation and worst-case prediction when compared with UNI and IBOSS subsamples with $n=2000$, which again demonstrates the significant savings achieved by a balanced subsample in observing the response.

\section{Real data application}
We consider the application to an online store offering clothing for pregnant women. The data are from five months of 2008 and include, among others, product category (4 levels), product code (217 levels), color (14 levels), model photography (2 levels), location of the photo on the page (6 levels), page number (5 levels), country of origin of the IP address of customers clicking the page (47 levels), month (5 levels), and product price in US dollars (continuous). The data contain more predictors to study the behavior patterns of customers. We are only using the above predictors to predict the product price and demonstrate the superiority of balanced subsamples. Further information on the dataset can be found in \cite{lapczynski2013discovering}.

\begin{figure}[t!]
  \centering
  \includegraphics[width=.7\textwidth]{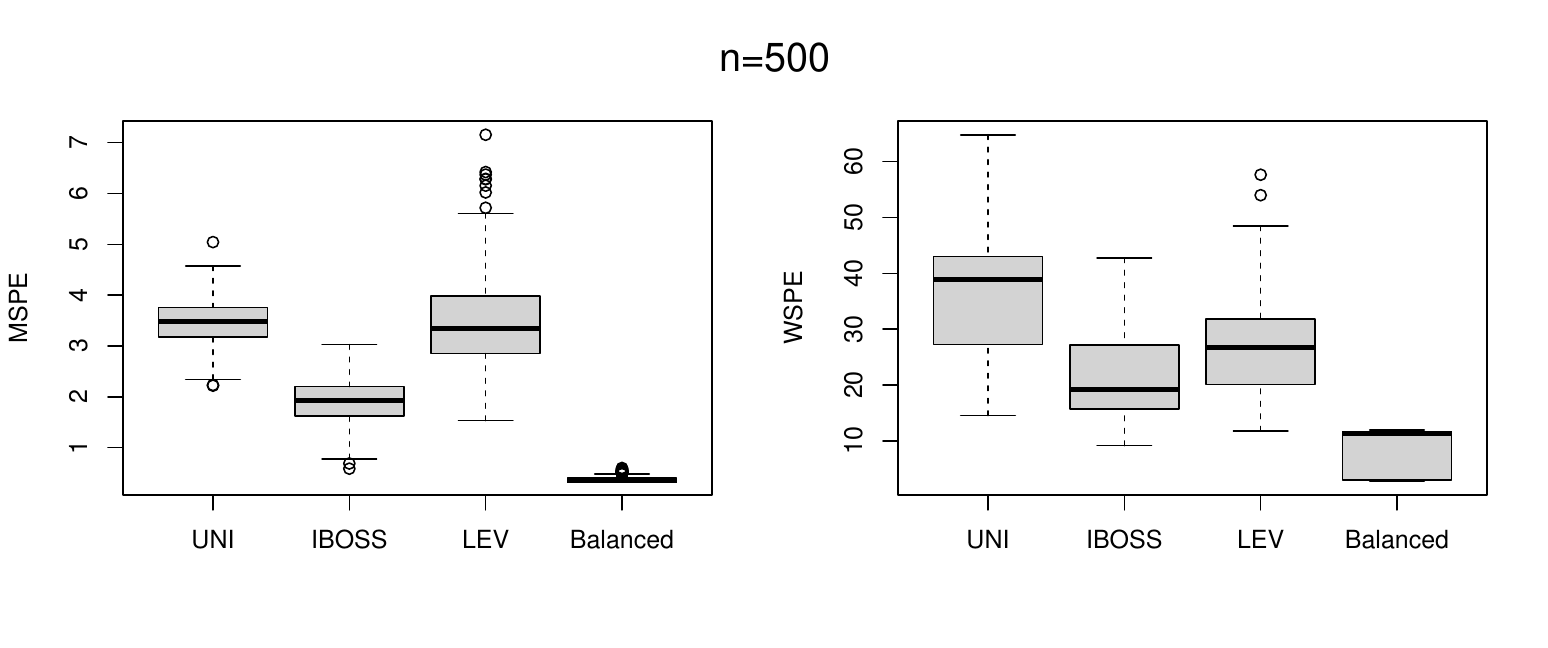}
  \includegraphics[width=.7\textwidth]{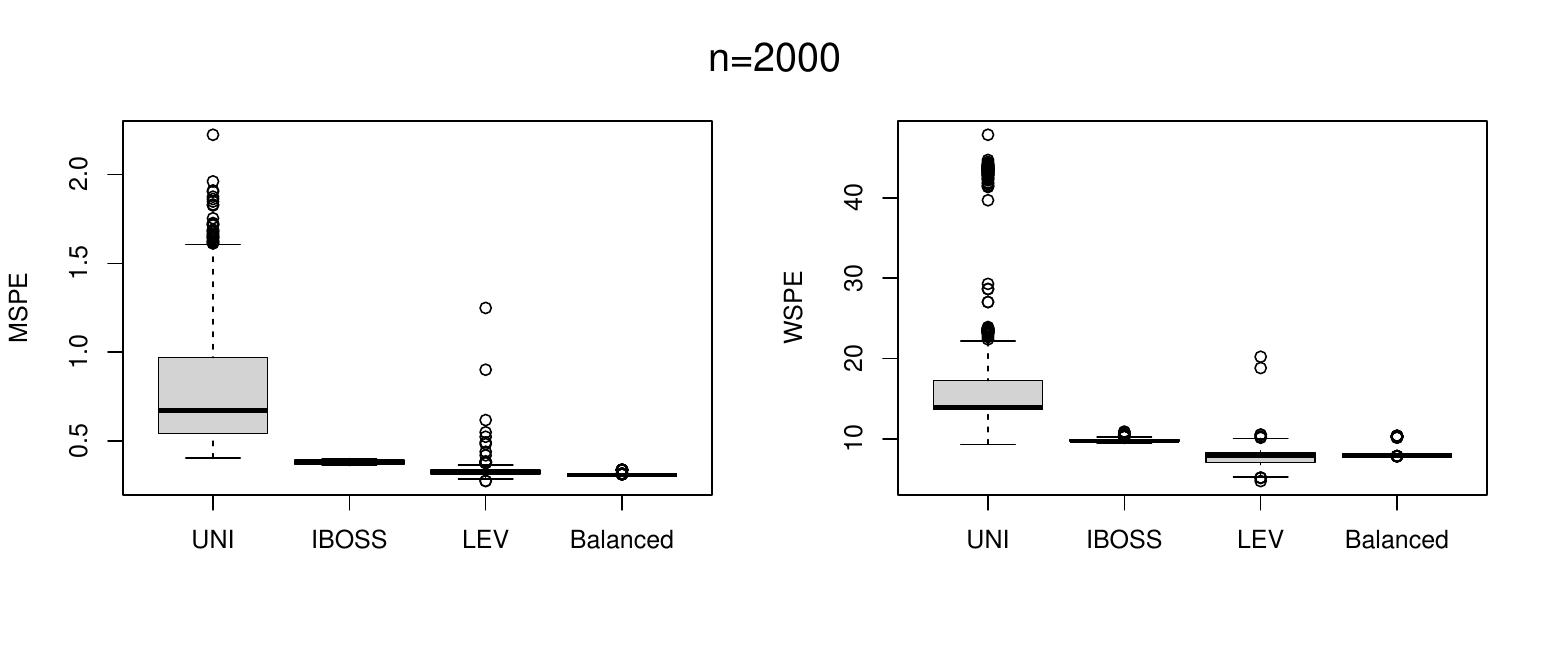}
  \vspace{-8mm}
  \caption{The MSPEs and WSPEs over the full data for different subsampling methods with $n=500$ (top) and 2000 (bottom). }\label{LassoPred}
\end{figure}

The full dataset has $n=165,474$ data points. All predictors are categorical and are coded to dummy variables, which results in 293 dummy variables in total (intercept included). The response variable, product price, is accessible for the full data, facilitating the observation of responses for each subsample and allowing for comparison of their predictive performance.
We consider subsample sizes $n=500$ and $2000$ and select subsamples from the data with different methods. Considering that some of the dummy variables may not be significant in real-world data, we use LASSO regression \citep{tibshirani1996regression} to select important ones and train a predictive model. On each subsample, a LASSO regression model is trained via the R package ``glmnet'' \citep{friedman2010regularization} with the parameter $\lambda$ selected by cross-validation. The trained model is then used to predict the product price over the full data. We expect that a balanced subsample should outperform other subsamples in a penalized regression due to its linear nature. To investigate this,
we do 500 repetitions of this process and plot the MSPE (mean squared prediction error) and WSPE of the trained models over the full data in Figure \ref{LassoPred}.
Balanced subsamples consistently yield superior predictions compared to other subsamples. In particular, UNI subsamples exhibit the poorest predictive performance of the full dataset. Their selection of markedly different points and subsequent divergent predictions result in unstable performance. In contrast, IBOSS and LEV subsamples demonstrate greater stability and improved performance due to their more balanced predictors. Overall, balanced subsamples consistently outperform others across both settings of subsample size.

\section{Discussion}

In this paper, we propose the balanced subsampling method for big data with categorical predictors. The selected subsample achieves a balance among predictor levels, maximizing the overall information provided by the subsample.
A balanced subsample is typically nonsingular and allows more accurate parameter estimation and prediction. Simulations and a real-world application confirm the improved performance of the subsample selected by the balanced subsampling over other available subsampling methods. 
A balanced subsample offers substantial cost savings when observing the response. 
Although this paper assumes binary dummy variables for coding the categorical predictors, all theoretical results work if any nonsingular coding system (for example, an orthogonal coding framework such as the orthogonal polynomial) is used for coding the predictors. The superiority of a balanced subsample does not depend on the coding system. 

\reva{We adopt Algorithm \ref{alg1} to minimize $f(X_s)$ (alternatively, $\sum_{1\leq i<l\leq n}[\delta(x_i^*,x_l^*)]^2$) due to its straightforward implementation and efficiency. A primary drawback of Algorithm \ref{alg1} is its lack of guaranteed optimization of $f(X_s)$. If an optimal solution is imperative, various optimization algorithms, such as the simulated annealing algorithm \citep{Morris1995}, can be employed. Convergence is assured for these algorithms with sufficient iterations. Nonetheless, when dealing with extensive datasets, computational challenges may impede the efficacy of such algorithms. While the current Algorithm \ref{alg1} does not guarantee the minimization of $f(X_s)$, our extensive numerical results are strong evidence that the algorithm tends to efficiently produce a dramatically improved subsample relative to those found by other methods.}

The balanced subsampling can be combined with robust regression, such as the S estimator \citep{rousseeuw1984robust}, to enhance the robustness of the trained model to possible outliers. The training of the estimator involves repeatedly selecting small and nonsingular subsamples from the full data, which, as discussed in this paper and \citep{koller2017nonsingular}, is infeasible via simple random sampling. The randomness and nonsingularity of balanced subsamples make them applicable to training such estimators, although their performance for this purpose requires further study.

Though the proposed balanced subsampling is focused on big data with categorical predictors, it can be modified and generalized to select subsamples with numerical or mixed-type predictors. To do so, we may discretize the numerical predictors and apply Algorithm \ref{alg1} to the discretized data. The selected subsample will cover the region of the full data evenly and uniformly and, therefore, promote a fair study and robust prediction over the region. 
\section*{Supplementary Materials}

The online supplementary materials provide proof of the theoretical results and discuss the computational complexity of the proposed algorithm.
\par
\section*{Acknowledgements}

The author gratefully acknowledges the support from the Central Indiana Corporate Partnership AnalytiXIN Initiative.
\par


\bibliographystyle{chicago}
\bibliography{ref}

\vskip .65cm
\noindent
Lin Wang\\
Department of Statistics\\ Purdue University, West Lafayette, IN 47907, USA
\vskip 2pt
\noindent
E-mail: linwang@purdue.edu
\vskip 2pt

\end{document}



\renewcommand{\baselinestretch}{2}

\markright{ \hbox{\footnotesize\rm Statistica Sinica: Supplement
}\hfill\\[-13pt]
\hbox{\footnotesize\rm
}\hfill }

\markboth{\hfill{\footnotesize\rm LIN WANG} \hfill}
{\hfill {\footnotesize\rm BALANCED SUBSAMPLING} \hfill}

\renewcommand{\thefootnote}{}
$\ $\par \fontsize{12}{14pt plus.8pt minus .6pt}\selectfont


 \centerline{\large\bf BALANCED SUBSAMPLING FOR BIG DATA}
\vspace{2pt}
 \centerline{\large\bf WITH CATEGORICAL PREDICTORS}
\vspace{.25cm}
 \author{Lin Wang}
\vspace{.4cm}
 \centerline{\it Lin Wang} 
 \centerline{\it Purdue University}
\vspace{.55cm}
 \centerline{\bf Supplementary Material}
\vspace{.55cm}
\fontsize{9}{11.5pt plus.8pt minus .6pt}\selectfont
\noindent

\setcounter{section}{0}
\setcounter{equation}{0}
\setcounter{table}{0}
\def\theequation{S\arabic{section}.\arabic{equation}}
\def\thesection{S\arabic{section}}
\def\thetable{S\arabic{section}.\arabic{table}}

\fontsize{12}{14pt plus.8pt minus .6pt}\selectfont

\section{Notations and Orthonormal Contrasts}
Recall that $\mathcal{X}$ denotes the set of all possible level combinations of predictors, that is, $\mathcal{X}=\{x=(x_1,\ldots,x_p): x_j=1,\ldots,q_j, j=1,\ldots, p\}$. Let $\mathcal{N}=\#\mathcal{X}=\prod_{j=1}^{p}q_j$.
Let $\mathcal{Z}$ be the matrix of dummy variables for $\mathcal{X}$ and $C$ be the coded matrix for $\mathcal{X}$ via orthonormal contrasts \citep{chen2022study,wang2022AClass} with $C^TC=\mathcal{N}I$, where $I$ is a conformable identity matrix. Then there exists a transformation matrix $P$ such that $\mathcal{Z}=CP$. Because both $\mathcal{Z}$ and $C$ have full column ranks, so $P$ is nonsingular. Clearly, rows of $Z_s$ come from rows of $\mathcal{Z}$, so $Z_s=C_sP$ and $M_s=P^TC_s^TC_sP$, where rows of $C_s$ are from the corresponding rows of $C$. 

\reva {Common orthonormal contrasts include (normalized) Helmert contrasts \citep{chambers2017statistical}, orthogonal polynomial contrasts \citep{wang2022AClass}, and complex contrasts \cite{xu2001generalized}. For example, the Helmert contrasts are used to contrast the second level with the first, the third with the average of the first two, and so on. For the $j$th predictor with $q_j$ levels, the $l$th contrast ($l=1,\ldots,q_j-1$) at level $u$ is 
\begin{equation}\label{helmert}
    c_{jl}(u) = \sqrt{\frac{q_j}{l+l^2}}\cdot a_{jl}(u),
    \text{ where }
    a_{jl}(u)=\begin{cases}
        -1, & \text{for $u<l$}\\
        l, & \text{for $u=l$}\\
            0, & \text{otherwise}
    \end{cases}
\end{equation}
Below is an example of the $a_{jl}(u)$ for a predictor with $q_j=5$ levels, where the $l$th column lists the $l$th contrast at $u=1,\ldots,5$ levels: 
$$
\begin{pmatrix}
-1 &  -1 &  -1  & -1 \\
1  & -1 &  -1  & -1 \\
0  &  2  & -1 &  -1 \\
0  &  0  &  3  & -1 \\
0  &  0  &  0  &  4
\end{pmatrix}.
$$
}

\section{Proof of Theorem 1}
\setcounter{equation}{0}
Since $M_s=P^TC_s^TC_sP$, then
\begin{equation}\label{eq2}
\lambda_{\min}(M_s)=\lambda_{\min}(P^TC_s^TC_sP)\geq \lambda_{\min}(P^TP)\lambda_{\min}(C_s^TC_s)=\nu \lambda_{\min}(C_s^TC_s),
\end{equation}
where $\nu=\lambda_{\min}(P^TP)>0$.
Let $\tilde{C}=I-C_s^TC_s/n=C^TC/\mathcal{N}-C_s^TC_s/n$. \reva{Because each level $u$ for the $j$th predictor appears $\mathcal{N}/q_j$ times in $\mathcal{X}$ and $n_j(u)$ times in $X_s$, }
then
$$
\tilde{C}=\left(
                           \begin{array}{ccccc}
                             0 & C_1^T & C_2^T & \cdots & C_p^T \\
                             C_1 & C_{11} & C_{12}^T & \cdots & C_{1p}^{T} \\
                             C_2 & C_{12} & C_{22} & \cdots & C_{2p}^T \\
                             \vdots & \vdots & \vdots & \ddots & \vdots \\
                             C_p & C_{1p} & C_{2p} & \cdots & C_{pp} \\
                           \end{array}
                         \right)
$$
where $C_j$ is a $(q_j-1)$-vector with the $l$th entry $\sum_{u=1}^{q_j}[q_j^{-1}-n_{j}(u)/n]c_{jl}(u)=\sum_{u=1}^{q_j}[q_j^{-1}-n_{j}(u)/n]c_{jl}(u)c_{j0}(u)$ with $c_{j0}(u)=1$, $C_{jj}$ is a $(q_j-1)\times (q_j-1)$ matrix with the $(l,m)$th entry $\sum_{u=1}^{q_j}[q_j^{-1}-n_{j}(u)/n]c_{jl}(u)c_{jm}(u)$, $C_{jk}$ is a $(q_j-1)\times (q_k-1)$ matrix with the $(l,m)$th entry $\sum_{u=1}^{q_j}\sum_{v=1}^{q_k}[(q_jq_k)^{-1}-n_{jk}(u,v)/n]c_{jl}(u)c_{km}(v)$, and
$c_{jl}(u)$ is the coded value for the $l$th contrast of the $j$th predictor at level $u$.
\reva{For example, for the Helmert contrast, $c_{jl}(u)$ is given in \eqref{helmert}}.
We have
$$
\lambda_{\max}(\tilde{C}) \leq \|\tilde{C}\|_F =\sqrt{f_1+f_2},
$$
where $\|\cdot\|_F$ denotes the Frobenius norm and
\begin{eqnarray*}
f_1&=&\sum_{j=1}^{p}\sum_{l=0}^{q_j-1}\sum_{m=0}^{q_j-1}\sum_{u=1}^{q_j}\left[\frac{1}{q_{j}}-\frac{n_{j}(u)}{n}\right]^2c_{jl}^2(u)c_{jm}^2(u)\\
f_2&=&\sum_{j=1}^{p}\sum_{k=1,k\neq j}^{p}\sum_{l=1}^{q_j-1}\sum_{m=1}^{q_k-1}\sum_{u=1}^{q_j}\sum_{v=1}^{q_k}\left[\frac{1}{q_jq_k}-\frac{n_{jk}(u,v)}{n}\right]^2c_{jl}^2(u)c_{km}^2(v).
\end{eqnarray*}
Because $\sum_{l=0}^{q_j-1}\sum_{m=0}^{q_j-1}c_{jl}^2(u)c_{jm}^2(u)\leq q_j^2$ and $\sum_{l=1}^{q_j-1}\sum_{m=1}^{q_k-1}c_{jl}^2(u)c_{km}^2(v)\leq q_jq_k$ for any $j,k$ and $u, v$, then
\begin{eqnarray*}
\lambda_{\max}(\tilde{C})
  &\leq& \sqrt{\sum_{j=1}^{p}\sum_{u=1}^{q_{j}}q_{j}^2\left[\frac{1}{q_{j}}-\frac{n_{j}(u)}{n}\right]^2
+\sum_{j=1}^{p}\sum_{k=1,k\neq j}^{p}\sum_{u=1}^{q_{j}}\sum_{v=1}^{q_{k}}q_jq_k\left[\frac{1}{q_jq_k}-\frac{n_{jk}(u,v)}{n}\right]^2}\\
&=& f(X_s).
\end{eqnarray*}
Since $C_s^TC_s=n(I-\tilde{C})$, $\lambda_{\min}(C_s^TC_s)= n (1-\lambda_{\max}(\tilde{C}))\geq n(1-f(X_s))$. Then by \eqref{eq2}, $\lambda_{\min}(M_s)\geq n\nu(1-f(X_s)).$

\section{Proof of Theorem 2}
\setcounter{equation}{0}

Since $M_s=P^TC_s^TC_sP$, where $P$ is a $Q\times Q$ transformation matrix with $Q=1+\sum_{j=1}^{p}(q_j-1)$, then
$$
\det(M_s)=\det(P^TC_s^TC_sP)=\det(P)^2\det(C_s^TC_s).
$$
Because $P$ is independent from the selection of $Z_s$, we only need to consider $\det(C_s^TC_s)$:
\begin{eqnarray}
  \det(C_s^TC_s) &=& \prod_{j=1}^{Q}\lambda_{j}(C_s^TC_s) \nonumber\\
   &\leq& \left\{\frac{\sum_{j=1}^{Q}\lambda_{j}(C_s^TC_s)}{Q}\right\}^{Q}  \nonumber\\
   &=& \left\{\frac{\tr(C_s^TC_s)}{Q}\right\}^{Q} \label{app1}\\
   &=& \left\{\frac{nQ}{Q}\right\}^{Q}\label{app2}\\
   &=& n^{Q}, \nonumber
\end{eqnarray}
where $\lambda_{j}(C_s^TC_s)$'s for $j=0,1,\ldots,Q$ are eigenvalues of $C_s^TC_s$, and $\tr(C_s^TC_s)=\tr(C_sC_s^T)=nQ$ because rows of $C_s$ are orthonormal. If $f(X_s)=0$, $X_s$ forms an orthogonal array and $C_s^TC_s=nI$, then $\det(C_s^TC_s)=n^{Q}$. This completes the proof.

\section{Proof of Theorem 3}
\setcounter{equation}{0}
We have
$$
E[(Y-z^T\hat{\beta}_s)^2|X_s]=E[(Y-z^T\beta)^2]+E[(z^T\beta-z^T\hat{\beta}_s)^2|X_s]=\sigma^2(1+z^TM_s^{-1}z)
$$
and
\begin{eqnarray*}
  \sum_{x\in \mathcal{X}}z^TM_s^{-1}z &=& \tr(\mathcal{Z}M_s^{-1}\mathcal{Z}^T)=\tr\{\mathcal{Z^TZ}M_s^{-1}\}=\tr\{P^TC^TCP(P^TC_s^TC_sP)^{-1}\} \\
  &=& \tr\{C^TC(C_s^TC_s)^{-1}\}=\mathcal{N}\tr\{(C_s^TC_s)^{-1}\} \geq \mathcal{N}Q^2/\tr(C_s^TC_s) = \mathcal{N}Q/n,
\end{eqnarray*}
where the last equation holds because $\tr(C_s^TC_s)=nQ$ following \eqref{app1} and \eqref{app2}.
Therefore, $\max_{x\in \mathcal{X}}z^TM_s^{-1}z\geq Q/n$ and
$\max_{x\in \mathcal{X}}E[(Y-z^T\hat{\beta}_s)^2|X_s]\geq \sigma^2(1+Q/n).$
On the other hand, when $f(X_s)=0$, $X_s$ is balanced, and then for any $z$, $z^TM_s^{-1}z=z^TP^TPz/n=\|Pz\|_2^2/n$. Note that $Pz$ is a row vector of $C$. The sum of squares of the $i$th row of $C$ is
$(1+\sum_{j=1}^{p}\sum_{l=1}^{q_j-1}c_{i,jl}^2)/n=(1+\sum_{j=1}^{p}(q_j-1))/n=Q/n$. Therefore, $z^TM_s^{-1}z=Q/n$ and $E[(Y-z^T\hat{\beta}_s)^2|X_s]=\sigma^2(1+Q/n).$ This completes the proof.


\section{Proof of Theorem 4}
\setcounter{equation}{0}
For a subsample $X_s=(x_{ij}^*)$, it can be verified that $\sum_{u=1}^{q_j}n_j(u)=n$, $\sum_{u=1}^{q_j}\sum_{v=1}^{q_k}n_{jk}(u,v)=n$, and $\sum_{i=1}^{n}\sum_{l=1}^{n}\delta_1(x_{ij}^*,x_{lj}^*)\delta_1(x_{ik}^*,x_{lk}^*)=\sum_{u=1}^{q_j}\sum_{v=1}^{q_k}n_{jk}(u,v)^2$ for any $j,k=1,\ldots,p.$
Then
\begin{eqnarray*}
  &&f^2(X_s) \\
  &=& \sum_{j=1}^{p}\sum_{u=1}^{q_j}\left(1-\frac{2q_jn_j(u)}{n}+\frac{q_j^2n_j(u)^2}{n^2}\right) + \sum_{j=1}^{p}\sum_{k=1,k\neq j}^{p}\sum_{u=1}^{q_{j}}\sum_{v=1}^{q_{k}}\left(\frac{1}{q_jq_k}-\frac{2n_{jk}(u,v)}{n}+\frac{q_jq_kn_{jk}(u,v)^2}{n^2}\right)\\
   &=& -\sum_{j=1}^{p}q_j+n^{-2}\sum_{j=1}^{p}q_j^2\left[\sum_{u=1}^{q_j}n_j(u)^2\right]-p(p-1)+ n^{-2}\sum_{j=1}^{p}\sum_{k=1,k\neq j}^{p}q_jq_k\left[\sum_{u=1}^{q_{j}}\sum_{v=1}^{q_{k}}n_{jk}(u,v)^2\right]\\
   &=& n^{-2}\sum_{j=1}^{p}\sum_{k=1}^{p}q_jq_k\left[\sum_{u=1}^{q_j}\sum_{v=1}^{q_k}n_{jk}(u,v)^2 \right]-\sum_{j=1}^{p}q_j-p(p-1)\\
   &=& n^{-2}\sum_{i=1}^{n}\sum_{l=1}^{n}\sum_{j=1}^{p}\sum_{k=1}^{p}q_jq_k\delta_1(x_{ij}^*,x_{lj}^*)\delta_1(x_{ik}^*,x_{lk}^*)-\sum_{j=1}^{p}q_j-p(p-1)\\
   &=& n^{-2}\sum_{i=1}^{n}\sum_{l=1}^{n}[\delta(x_i^*,x_l^*)]^2-\sum_{j=1}^{p}q_j-p(p-1)\\
   &=& 2n^{-2}\sum_{1\leq i<l\leq n}[\delta(x_i^*,x_l^*)]^2+C.
\end{eqnarray*}


\section{Computing complexity and time}

The computational complexity of Algorithm 1 is $O(Npn)$. The computational complexity of IBOSS is $O(N(\sum_{j=1}^{p}q_j))=O(Np \bar{q})$, where $\bar{q}$ represents the average number of levels of predictors. For LEV, we use a fast Singular Value Decomposition method implemented in the R package ``corpcor'' to accelerate LEV, so the complexity is also $O(N(\sum_{j=1}^{p}q_j))=O(Np \bar{q})$. 

Tables \ref{tab1} and \ref{tab2} show the computing time for the subsampling process in the setting of the simulation studies. All computations are carried out on a laptop running Windows 11 Pro with an Intel Core i7-12700H processor and 32GB memory. 
When $n=500$, the running time of balanced subsampling is comparable to that of fast LEV, with both being relatively faster than IBOSS. When $n=2000$, the running time of balanced subsampling increases to four times that of $n=500$, making its running time around four times that of fast LEV and 2.5 times that of IBOSS. 

Given the constraints of limited resources for labeling data (observing the response), the subsample size $n$ is typically not large. In such scenarios, balanced subsampling exhibits comparable computational time to other subsampling methods, making it viable for handling large datasets.

Table \ref{tab3} presents the computing times for the real data application. With the real data featuring more levels and a larger $\bar{q}$, the advantage of balanced subsampling in terms of computing time is evident.

However, the primary objective of this paper is to identify an optimal subsampling approach for measurement-constrained regression, with the goal of achieving superior estimation and predictive performance. While computational efficiency is undoubtedly important, it takes a secondary role to the primary concern of achieving optimal performance.

\begin{table}
    \centering
    \caption{Running time (in seconds) of subsampling methods when $p=20$, $q=2,\ldots,(p+1)$, and the subsample size is $n=500$.}
    \smallskip
    \begin{tabular}{|ccccc|}
    \hline
        $N$ & UNI & IBOSS & LEV & Balanced\\
        \hline
        $10^4$ & 0 & 1.63 & 1.09 & 1.19 \\
        $10^5$ & 0 & 20.51  & 11.95 & 13.88 \\
        $10^6$ & 0 & 207.63  & 124.20 & 131.40 \\
        \hline
    \end{tabular}
    \label{tab1}

    \centering
    \caption{Running time (in seconds) of subsampling methods when $p=20$, $q=2,\ldots,(p+1)$, and the subsample size is $n=2000$.}
    \smallskip
    \begin{tabular}{|ccccc|}
    \hline
        $N$ & UNI & IBOSS & LEV & Balanced\\
        \hline
        $10^4$ & $4\times 10^{-4}$ & 1.56 & 1.05 & 4.21 \\
        $10^5$ & $8\times 10^{-4}$ & 20.96  & 12.68 & 56.31 \\
        $10^6$ & $0.002$ & 228.88  & 131.90 & 563.68 \\
        \hline
    \end{tabular}
    \label{tab2}
    
    \centering
    \caption{Running time (in seconds) of the subsampling process for the real data.}
    \smallskip
    \begin{tabular}{|ccccc|}
    \hline
        $n$ & UNI & IBOSS & LEV & Balanced\\
        \hline
        $500$ & $4\times 10^{-4}$ & 26.39 & 26.34 & 4.24 \\
        $2000$ & $0.0012$ & 40.53  & 31.41 & 21.34 \\
        \hline
    \end{tabular}
    \label{tab3}
\end{table}









\bibliographystyle{chicago}
\bibliography{ref}